\documentclass[aps,prf,superscriptaddress,groupedaddress]{revtex4}
\usepackage{graphicx,color}  
\usepackage{bm}        % for math
\usepackage{amssymb}   % for math
\usepackage{gensymb}
\usepackage{comment} 
\usepackage[normalem]{ulem} % for visible strike-through revisions
\usepackage[colorlinks, citecolor=red]{hyperref}

\begin{document}

%\title{Jet coronation: Impact-induced jets wearing a crown via cavitation dynamics}
\title{Jet coronation: Coexistence of compressible and incompressible dynamics}
\author{H. Watanabe}
\author{K. Hashimoto}
\author{W. K. A. Worby}
\affiliation{Department of Mechanical Systems Engineering, Tokyo University of Agriculture and Technology, Nakacho 2-24-16 Koganei, Tokyo 184-8588, Japan}
\author{A. Kiyama}
\affiliation{
Graduate School of Science and Engineering, Saitama University, Shimo-okubo 255, Sakura-ku, Saitama-shi, Saitama 338-8570, Japan
}
\author{L. Kahouadji}
\author{O. K. Matar}
\affiliation{
Department of Chemical Engineering, Imperial College London, London SW7 2AZ, United Kingdom
}
\author{Y. Tagawa}
%\email{tagawayo@cc.tuat.ac.jp}
\affiliation{Department of Mechanical Systems Engineering, Tokyo University of Agriculture and Technology, Nakacho 2-24-16 Koganei, Tokyo 184-8588, Japan}

\begin{abstract}
This paper is associated with a poster winner of a 2025 American Physical Society’s Division of Fluid Dynamics (DFD) Gallery of Fluid Motion Award for work presented at the DFD Gallery of Fluid Motion. The original poster is available online at the Gallery of Fluid Motion, https://gfm.aps.org/meetings/dfd-2025/692e39e5a7f805227b16fe9e
\end{abstract}

\maketitle

\begin{figure}[h]
\centering
\includegraphics[width=0.95\linewidth]{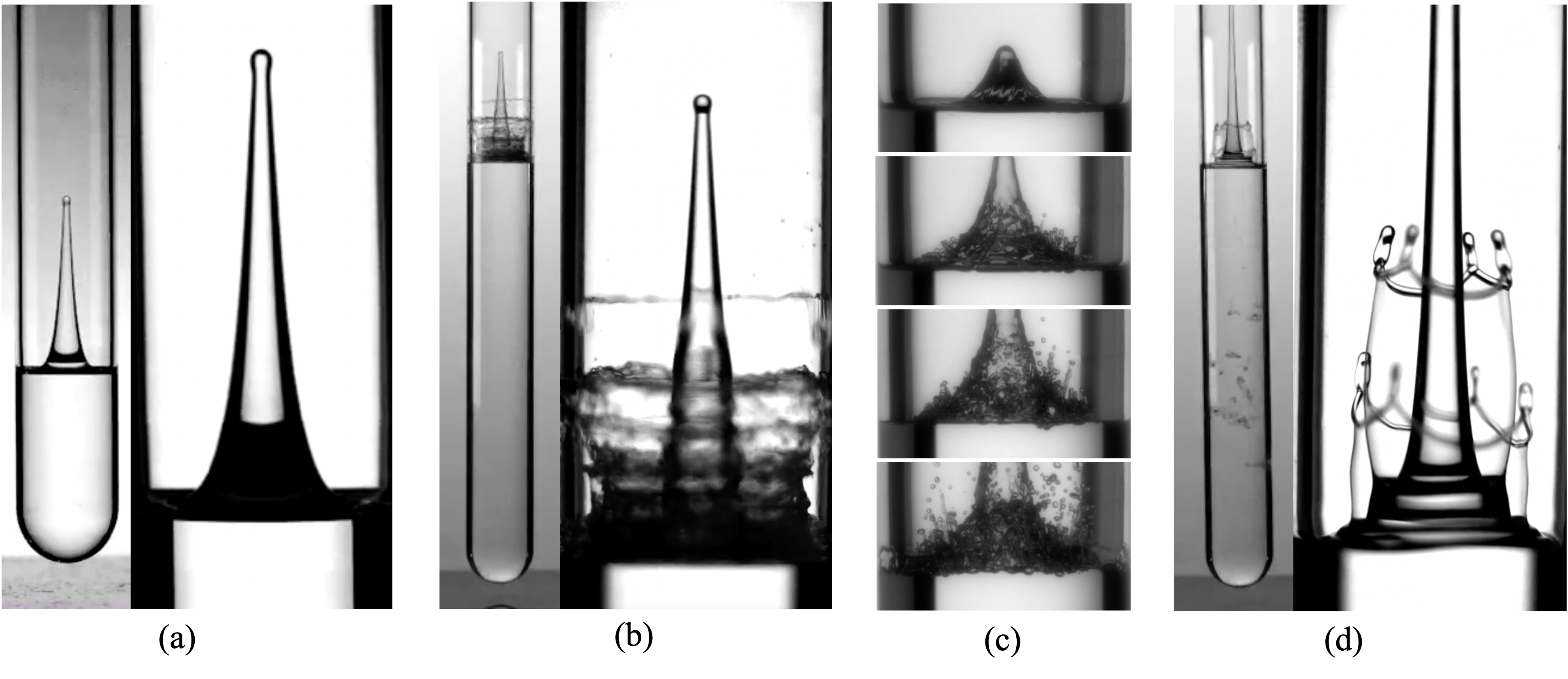}
\caption{\label{fig:Intro} High-speed visualisation of the different jetting regimes generated during the impact of partially filled test tubes \cite{Kiyama_jfm_2016} and a flask \cite{Watanabe_ijmf_2025} on a solid substrate. (a) Formation of a regular, smooth, and axisymmetric liquid jet emerging from the free surface. (b) A jet surrounded by outward-rising crowns produced by interfacial instabilities and radial liquid ejection near the jet base. (c) Progressive destabilisation and atomisation into fine droplets caused by the combined dynamics of the jet and surface waves observed during flask impact. (d) A crown-jet configuration characterised by vapour cavities simultaneously generated within the liquid bulk.}
\end{figure}

The impact of a liquid-filled container on a rigid substrate gives rise to a rich spectrum of transient hydrodynamic phenomena, including jet focusing, crown formation, interfacial-instability-induced atomisation, and cavitation. Upon impact, the abrupt change in the container velocity impulsively accelerates the liquid column, producing a vertically directed focused jet at the free surface. Depending on the impact conditions, container geometry, and the physical properties of the air--liquid system, the jet may either remain smooth and axisymmetric or evolve into more complex dynamics characterised by surface-wave development, droplet ejection, and, perhaps most remarkably, the formation of vapour cavities.

The dynamics of impact-induced focused jets have attracted sustained attention because of their relevance to a broad range of natural and industrial processes involving impulsive fluid motion, cavitation physics, and high-speed jet generation. The generation of a jet from the centre of the air--liquid interface induced by container impact can be traced back to the reports of Milgram \cite{Milgram_jfm_1969} and Lavrentiev and Chabat \cite{Lavrentiev_Book_1980}, and is widely known as \textit{\textbf{Pokrovski's experiment} }\cite{Lavrentiev_Book_1980}. Building upon these foundations, Kiyama \textit{et al.} \cite{Kiyama_jfm_2016} demonstrated the critical influence of water-hammer effects and cavitation during jet formation in partially filled tubes impacting a rigid substrate, thereby highlighting the essential role of liquid compressibility and acoustic-wave interactions in governing jet dynamics. Related studies have further clarified the physical ingredients that underlie this problem. Pan \textit{et al.} \cite{Pan_pnas_2017} identified acceleration-induced cavitation as a distinct mechanism in impulsively accelerated liquids, while Kurihara \textit{et al.} \cite{Kurihara_jfm_2025} showed that pressure fluctuations under short-time acceleration can be organised by the ratio between the acoustic time scale and the acceleration duration. In the complementary incompressible limit, Gordillo \textit{et al.} \cite{Gordillo_jfm_2020} described impulsive jet generation by flow focusing based on the initial interfacial velocity field. These studies motivate the present view that impact-induced jets can involve both pressure-impulse-driven, nearly incompressible focusing and local acoustic/cavitation dynamics. Subsequently, Yukisada \textit{et al.} \cite{Yukisada_Langmuir_2018} showed that pre-existing surface bubbles on the inner wall of the container can increase the focused jet velocity by significantly reducing the peak negative pressure of the pressure wave. More recently, Watanabe \textit{et al.} \cite{Watanabe_ijmf_2025} investigated the effect of converging container geometries on impact-induced jet focusing, demonstrating that geometric confinement can significantly amplify jet velocity while profoundly modifying the resulting interfacial dynamics.

Figure \ref{fig:Intro} illustrates several characteristic jetting regimes arising from the impact of a partially filled test tube or flask on a rigid substrate, revealing the rich spectrum of interfacial dynamics that can emerge under different impact conditions and liquid properties. Under relatively weak impacts, the free surface gives rise to a regular, slender, and highly focused liquid jet that ascends along the central axis while preserving an almost perfectly axisymmetric form, as shown in Fig. \ref{fig:Intro}(a). In this regime, the jet remains smooth and stable, indicating that inertial focusing dominates over destabilising capillary and aerodynamic effects during the early stages of the evolution.
As the impact intensity increases, significantly stronger disturbances develop at the air--liquid interface, giving rise to outward-expanding liquid crown structures surrounding the central jet, as illustrated in Fig. \ref{fig:Intro}(b). These crowns emerge from the radial ejection of liquid near the jet base and are accompanied by the amplification of interfacial instabilities along the expanding rim. The resulting morphology consists of a highly focused central jet enclosed by thin annular liquid sheets, which progressively deform in the radial direction as they propagate upward toward the tube wall. 
Under different initial conditions, a more complex dynamical regime emerges through the coupling between the focused jet motion and Faraday-type surface waves generated within the liquid column. As shown in Fig. \ref{fig:Intro}(c), the impact of a partially filled flask gives rise to oscillatory interfacial modes together with an upward jet, and the rapid destabilisation of the air--liquid interface eventually leads to atomisation into fine droplets \cite{Faraday_ptrs_1831,Vukasinovic_jfm_2007,Panda_prf_2023}. This regime is characterised by increasingly disordered surface structures, intense wave amplification, and fragmentation of the liquid interface.
Finally, Fig. \ref{fig:Intro}(d) presents a crown–jet configuration accompanied by cavitation phenomena. In this regime, vapour cavities form within the bulk liquid beneath the rising jet as a result of the intense pressure fluctuations generated during the impact event. The coexistence of focused jetting, crown formation, and cavitation underscores the strongly nonlinear nature of the flow and reveals the intricate interplay among inertial focusing, interfacial instabilities, wave dynamics, and pressure-driven cavity formation, together with the simultaneous presence of both compressible and incompressible regions throughout the system. Here, the central jet and the crown-like liquid sheets are associated primarily with inertial focusing and free-surface motion, whereas the appearance of vapour cavities and pressure fluctuations indicates a locally compressible response of the liquid column.

These studies collectively underline the complex interplay among impact-induced pressure waves, free-surface instabilities, cavitation, and geometric focusing mechanisms governing the formation and evolution of impulsive liquid jets. In the present study, we revisit the “cavitation-type jet” reported by Kiyama et al. \cite{Kiyama_jfm_2016}, with particular emphasis on the dynamics of the secondary sheet-like jet and the evolution of cavitation bubbles. Rather than re-establishing the basic water-hammer/cavitation mechanism, we highlight the secondary annular sheet and crown dynamics, their rim instabilities, and their apparent coupling with cavitation-bubble activity.

\begin{figure}[h]
\centering
\includegraphics[width=1\linewidth]{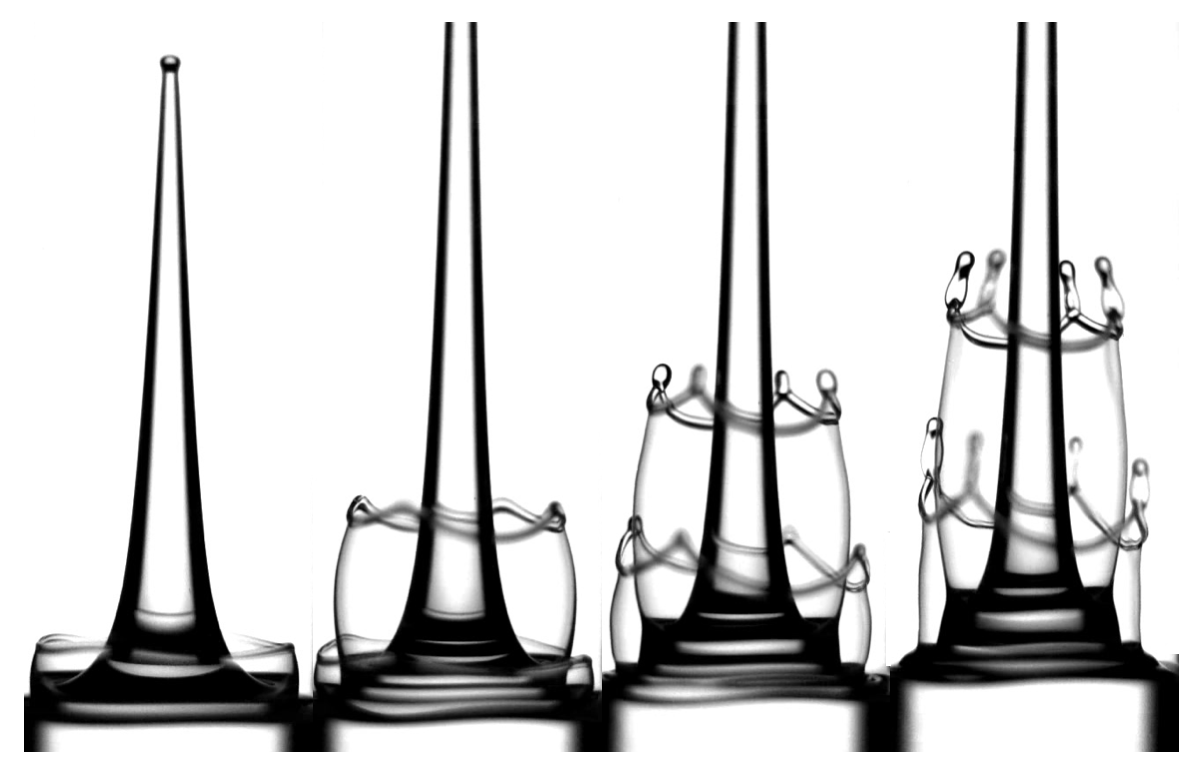}
\caption{\label{fig:crown} Formation of a focused liquid jet and crown-like sheet jets upon impact. The panels are shown, from left to right, at \(t = 3.2\), \(4.9\), \(6.6\), and \(8.2~\mathrm{ms}\), respectively, where \(t = 0~\mathrm{ms}\) denotes the moment of container impact. See Supplementary Movie~1 \cite{Supp1} for the corresponding animation.}
\end{figure}

The experimental configuration employed to investigate the crown-formation regime depicted in Figs.~\ref{fig:Intro}(d) and~\ref{fig:crown} consists of a cylindrical glass test tube (inner diameter \(D = 14.2~\mathrm{mm}\), wall thickness \(1.2~\mathrm{mm}\)) partially filled with silicone oil (Sigma-Aldrich, kinematic viscosity \(\nu = 10~\mathrm{mm}^2/\mathrm{s}\) at room temperature), whose surface tension is typically $\gamma \approx 20$--$21$ mN/m at room temperature. The tube is released from a prescribed height and allowed to undergo free fall before impacting a rigid substrate (Toshin Steel, SS400). Upon impact, the sudden deceleration of the container generates an impulsive pressure wave within the liquid column, leading to the formation of a focused jet at the air--liquid interface, along with complex interfacial dynamics such as crown formation, surface instabilities, and cavitation phenomena. The initial liquid height inside the tube was fixed at \(L = 103~\mathrm{mm}\), while the dropping height (the bottom of the test tube) was set to \(H = 135~\mathrm{mm}\), corresponding to a nominal impact velocity of \(U_0 \simeq \sqrt{2gH} = 1.63~\mathrm{m/s}\) when air resistance is neglected. These conditions were selected to produce sufficiently energetic impacts capable of generating cavitation-assisted jetting regimes. The rigid impact surface was designed to minimise energy dissipation and ensure reproducible impact conditions throughout the experiments.
To simultaneously visualise both the evolution of the air--liquid interface and the dynamics of the entire liquid column, two high-speed cameras (Photron FASTCAM SA-X) were synchronised during the experiments. Images were acquired at frame rates of \(30{,}000~\mathrm{fps}\) and \(20{,}000~\mathrm{fps}\), respectively, providing sufficient temporal resolution to capture the rapid jet formation and cavitation processes occurring immediately after impact. The two optical configurations provided spatial resolutions of  \(0.037~\mathrm{mm/pixel}\) and \(0.321~\mathrm{mm/pixel}\), respectively, enabling detailed observation of both the fine interfacial structures near the jet tip and the large-scale motion of the liquid column and cavitation bubbles. Further details regarding the experimental apparatus and optical alignment can be found in previous work~\cite{Kiyama_jfm_2016}.

The high-speed imaging results are presented in Fig.~\ref{fig:crown}. Immediately following the impact of the container (\(t = 0~\mathrm{ms}\)), a slender and highly focused liquid jet emerges from the initially concave air--liquid interface, as illustrated in the left panel of the figure. During the first few milliseconds following impact, the interface undergoes rapid deformation driven by inertial focusing of the surrounding liquid toward the central axis, resulting in the formation of an upward-moving jet. At approximately \(2.3~\mathrm{ms}\) after impact, a thin annular liquid sheet develops around the primary jet and expands upward, giving rise to an initially smooth, nearly axisymmetric liquid crown, as shown in the left panel corresponding to \(t = 3.2~\mathrm{ms}\).
As the flow evolves further, the crown rim continues to propagate upward while simultaneously slightly contracting toward the central axis, although no coalescence with the primary jet is observed during this stage. In parallel, small circumferential perturbations develop along the crown rim. These disturbances, which correspond to azimuthal modulations of the crown rim, appear to undergo progressive amplification in a manner reminiscent of the fluid-chain mechanism \cite{Hasha_pof_2002,Bush_jfm_2004,Kahouadji_prf_2025}, ultimately giving rise to the characteristic crown-like morphology and rim corrugation. The growth of these instabilities underscores the strong coupling among inertial forcing, capillary effects, and the evolving dynamics of the liquid sheet. The emergence of these fluid-chain-like structures may be associated with rim retraction combined with local mass conservation along the expanding liquid sheet. As the rim decelerates and begins to retract, perturbations are amplified through the redistribution of fluid along the circumference of the rim, promoting the development of azimuthal modulations. By contrast, this type of rim corrugation is much less pronounced when the annular sheet undergoes sustained radial expansion without significant rim retraction, as illustrated in Fig.~\ref{fig:Intro}(b), where the sheet dynamics remain comparatively smooth and quasi-stable.

The formation of an annular thin liquid sheet from the base of the air--liquid interface and the development of circumferential rim disturbances were repeatedly observed, as shown in Fig.~\ref{fig:crown} \((t = 4.9-8.2~\mathrm{ms})\). These phenomena occurred simultaneously with the quasi-periodic disappearance and reappearance of the vapour cavity generated beneath the interface (see Fig.~\ref{fig:Intro}(d)). Quantitative details of the temporal evolution of the total cavitation volume, the focused jet tip position, and the crown tip position are provided in Supplementary Movie~2 \cite{Supp2}, which further illustrates the coupled dynamics of cavitation and crown development over time. The total cavitation volume exhibited significant temporal fluctuations, and in particular, the high positive pressure expected to arise with bubble disappearance and reappearance \cite{Yukisada_Langmuir_2018} suggests a strong coupling between the pressure waves generated during bubble collapse and the formation of the crown from the base of the air--liquid interface.

This study visualises and highlights the diverse mechanisms that arise during the impact of a partially filled container on a rigid substrate. In particular, we focus on a regime in which crown formation, fluid-chain-like rim corrugation, and cavitation coexist, revealing a complex interplay between incompressible interfacial dynamics and locally compressible effects within the liquid column. The observed coupling among these phenomena highlights the highly nonlinear nature of the flow and underscores the richness of the underlying physics. In particular, a possible relationship is suggested between the crown structures shown in Fig.~\ref{fig:crown} and the growth and subsequent collapse of cavitation bubbles; however, the detailed mechanism remains unclear. Clarifying this coupling will require a more comprehensive investigation and constitutes an important direction for future work.

\vspace{\baselineskip}
This work was partly supported by Grant-in-Aid for JSPS Fellows Nos.
JP24KJ1023, JP25KJ1211, JSPS KAKENHI through Grant No. JP24H00289, the Japan
Science and Technology Agency PRESTO (Grant No. JPMJPR21O5), and
by funding from the Institute of Global Innovation Research at the
Tokyo University of Agriculture and Technology. O.K.M and L.K. acknowledge the support of the Engineering and Physical Sciences Research Council, UK, through the  PREMIERE (EP/T000414/1) programme grant, and the ANTENNA Prosperity Partnership grant (EP/V056891/1).

\vspace{\baselineskip}
\textit{Data availability.} The data supporting the findings of this study are available from the authors upon reasonable request.

%\begin{acknowledgments}
%\end{acknowledgments}

%\appendix
%\section{Appendixes}

%\bibliography{reference}% Produces the bibliography via BibTeX.

\end{document}